\documentclass[twocolumn,showpacs,preprintnumbers,amsmath,amssymb,superscriptaddress]{revtex4}
\usepackage{bm}
\usepackage{epsfig,amssymb,amsmath,bm}

\begin{document}

\title{On the description of vector bosons.}
\author{K. S. Karplyuk}
\email{ozhmudsky@physics.ucf.edu}
 \affiliation{Department of Radiophysics, Taras Shevchenko University, Academic 
Glushkov prospect 2, building 5, Kyiv 03122, Ukraine}

\begin{abstract}  
A new way of  describing massive
vector bosons is proposed. It is possible to change the boson
propagator in such way that the theory with the vector boson
interaction becomes a renormalized one. One example of the efficacy
of such a description is shown using the example of the weak
interaction of  charged currents.
\end{abstract}


\maketitle

The Proca equation is usually used  for the description of  massive vector bosons:
\begin{equation}
\partial_\mu F^{\mu\nu}+\varkappa^2X^\nu=\mathfrak{j}^\nu_e.
\end{equation}
Where,
$\varkappa=m/\hbar c$,
\begin{equation}
F^{\mu\nu}=\partial^\mu X^\nu-\partial^\nu X^\mu,
\end{equation}
$X^\nu$ is the vector potential, which is determined by the equation
\begin{equation}
\partial_\alpha \partial^\alpha X^\nu+\varkappa^2X^\nu=\mathfrak{j}^\nu_e+
\frac{1}{\varkappa^2}\partial^\nu\partial_\mu \mathfrak{j}^\mu_e.
\end{equation}
However,  Eqs. (1)-(3) lead to  unacceptable results. The
propagator of the boson described by  Eq. (1) has the form
\begin{equation}
D_{\mu\nu}=-\frac{\eta_{\mu\nu}}{k^2-\varkappa^2}+
\frac{1}{k^2-\varkappa^2}\frac{k_\mu k_\nu}{\varkappa^2},
\end{equation}
where $\eta_{\mu\nu}=\mathrm{diag}(1,-1,-1,-1)$, and $k_\mu$ is the
wave vector. The second term  on the right-hand side of Eq. (4) is
the offending term and makes  the theory with the boson interaction
a non-renormalized one. To save renormalization, in the
Weinberg-Salam electroweak model the contribution from the second
term is compensated for by the contribution from the hypothetical
Higgs bosons.

It should be noted that the second term in Eq. (4) is associated
with the non-conserved current in Eq. (3). If the current is
conserved, $\partial_\mu \mathfrak{j}^\mu_e=0$, and this term is absent.

We propose another way to eliminate the second term in Eq. (4). To describe
the vector bosons let us use the equation
\begin{equation}
\partial^\nu\epsilon+\partial_\mu F^{\mu\nu}+\varkappa^2X^\nu=
\mathfrak{j}^\nu_e
\end{equation}
instead of the Proca equation. Where the scalar field $\epsilon$ is given by:
\begin{equation}
\epsilon=\partial_\alpha X^\alpha.
\end{equation}
In this case the potential is given by the equation
\begin{equation}
\partial_\alpha \partial^\alpha X^\nu+\varkappa^2X^\nu=
\square X^\nu+\varkappa^2X^\nu=\mathfrak{j}^\nu_e,
\end{equation}
where $\square=c^{-2}\partial^2/\partial t^2-\Delta$ is the
d'Alambertian. The propagator of the boson described by equations
(5), and (7) has the form
\begin{equation} D_{\mu\nu}=-\frac{\eta_{\mu\nu}}{k^2-\varkappa^2}.
\end{equation}
Thus, the introduction of the scalar field $\epsilon$ improves
the propagator (4). The propagator (8) is only different
from the photon one by  the mass in the denominator. It allows
continuous transition to the massless boson, and coincides with the photon
propagator when $k\to \infty$.   The most important consequence is that the theory
based on such propagator becomes renormalized.

The necessity of introduction of the scalar field $\epsilon$ is
associated with a non-conserved current, but not associated with the
mass of the bosons. Differentiation of Eq. (5) yields the equation
for the scalar field $\epsilon$:
\begin{equation}
\partial_\nu\partial^\nu\epsilon+\varkappa^2\epsilon=\partial_\nu
\mathfrak{j}^\nu_e.
\end{equation}
As one can see the scalar field $\epsilon$ is generated only by a
non-conserved current, which does not satisfy the equation of continuity
\begin{equation}
\partial_\nu \mathfrak{j}^\nu_e=0.
\end{equation}

We must note that should the electric charge not be conserved, it will also be necessary
to change Maxwell's equations in a similar way.

The above consideration of the vector boson description can be
extended to the pseudovector bosons. To describe them we propose to use
the equation
\begin{equation}
\partial^\kappa\beta+\frac{1}{2}\varepsilon^{\kappa\lambda\mu\nu}
\partial_\lambda
F_{\mu\nu}+\varkappa^2Y^\kappa=\mathfrak{j}^\kappa_m,
\end{equation}
where
\begin{equation}
\beta=\partial_\nu Y^\nu,\hspace{7mm}
F^{\kappa\lambda}=\varepsilon^{\kappa\lambda\mu\nu}\partial_\mu
Y_\nu.
\end{equation}
Potential $Y^\nu$  is given by the same equation as potential
 $X^\nu$:
\begin{equation}
\partial_\alpha \partial^\alpha Y^\nu+\varkappa^2Y^\nu=
\square Y^\nu+\varkappa^2Y^\nu=\mathfrak{j}^\nu_m.
\end{equation}

Description of  vector and pseudovector bosons with the same mass
can be combined. It is useful to do, because the weak interaction is
just $V-A$ character.  We can introduce such a joint  description if we
associate the field $F^{\mu\nu}$ with the two potentials $X^\nu$ and $Y^\nu$
simultaneously,
\begin{equation}
F^{\kappa\lambda}=\partial^\kappa X^\lambda-\partial^\lambda
X^\kappa+\varepsilon^{\kappa\lambda\mu\nu}\partial_\mu Y_\nu,
\end{equation}
and use both the Eqs. (5) and (11). We write down
these equations in three-dimensional form:
\begin{align}
\frac{1}{c}\frac{\partial {\epsilon}}{\partial t}+\nabla\cdot{\bm
E}-\varkappa V^0&=\zeta c\rho_e,\\ \frac{1}{c}\frac{\partial {\bm
E}}{\partial t}-\nabla\times\bm{\mathfrak{B}}+ \nabla\epsilon
+\varkappa{\bm V}&=-\zeta{\bm j}_e,\\ -\frac{1}{c}\frac{\partial
{\beta}}{\partial
t}+\nabla\cdot\bm{\mathfrak{B}}+\varkappa U^0&=-\zeta c\rho_m,\\
\frac{1}{c}\frac{\partial\bm{\mathfrak{B}}}{\partial t}+\nabla\times
{\bm E}- \nabla\beta - \varkappa{\bm U}&=\zeta{\bm j}_m.
\end{align}
Here,
\begin{align}
{\bm E}&= -\frac{1}{c}\frac{\partial {\bm X}}{\partial
t}-\nabla X^0+\nabla\times\bm{Y},\\
\bm{\mathfrak{B}}&=\frac{1}{c}\frac{\partial\bm{Y}}{\partial
t}+\nabla Y^0+\nabla\times {\bm X},\\
\epsilon=\frac{1}{c}\frac{\partial X^0}{\partial
t}&+\nabla\cdot {\bm X},\hspace{0.5mm}V^0=-\varkappa X^0,
\hspace{0.5mm}{\bm V }=-\varkappa {\bm X},\\
\beta=\frac{1}{c}\frac{\partial Y^0}{\partial
t}&+\nabla\cdot\bm{Y},\hspace{0.5mm}U^0=-\varkappa
Y^0,\hspace{0.5mm} {\bm U }=-\varkappa\bm{Y}.
\end{align}
The three-dimensional coordinates of the vectors and tensors are
given by the rules: $\bm{E}=(F_{01},F_{02},F_{03})$,
$\bm{\mathfrak{B}}=(-F_{23},$ $-F_{31},-F_{12})$,
$\bm{X}=(X^1,X^2,X^3)$, $\bm{Y}=(Y^1,Y^2,Y^3)$, $\zeta c\rho_e=\zeta
j^0_e=\mathfrak{j}^0_e$, $\zeta\bm{j}_e=(\mathfrak{j}^1_e,
\mathfrak{j}^2_e,\mathfrak{j}^3_e)$, $\zeta c\rho_m=\zeta
j^0_m=\mathfrak{j}^0_m$, $\zeta\bm{j}_m=(\mathfrak{j}^1_m,
\mathfrak{j}^2_m,\mathfrak{j}^3_m)$. The coefficients
$c=1/\sqrt{\varepsilon_0\mu_0}$ and
$\zeta=\sqrt{\frac{\mu_0}{\varepsilon_0}}$ are introduced in the
definitions of the three-dimensional variables so that in case the
$\varkappa=0$, $j^\mu_m=0$, $Y^\mu=0$, $\epsilon=0$,  Eqs. (15)-(18)
coincide with Maxwell's equations, written down in SI unit system.
In this case we postulate $\bm{\mathfrak{B}}=c\bm{B}$,
$X^0=\varphi$, $\bm{X}=c\bm{A}$ and use the constants $c$ and
$\zeta$ instead of the constants $\varepsilon_0$ and $\mu_0$. This
form of  Maxwell's equations  allows us to use all the advantages of
the Gaussian unit system,  remaining at the same time in the SI unit
system.

The following equalities
\begin{align}
\frac{1}{c}\frac{\partial V^0}{\partial t}+\nabla\cdot{\bm
V}+\varkappa{\epsilon}&=0,\\ \frac{1}{c}\frac{\partial {\bm
V}}{\partial t}-\nabla\times {\bm U}+ \nabla V^0-\varkappa{\bm
E}&=0,\\ \frac{1}{c}\frac{\partial {\bm U}}{\partial t}+\nabla\times
{\bm V}+ \nabla U^0+
\varkappa\bm{\mathfrak{B}}&=0,\\
\frac{1}{c}\frac{\partial U^0}{\partial t}+\nabla\cdot{\bm U}+
\varkappa{\beta}&=0,
\end{align}
are satisfied due to the definitions (19)-(22). It may seem strange
to introduce the fields $V^\mu$ and $U^\mu$, which are different
from the potentials only by factor $\varkappa$. However, Eqs.
(15)-(18) and (23)-(26) allow the introduction of the tensor, scalar
and pseudoscalar potentials also. The fields $V^\mu$ and $U^\mu$ are
connected with such potentials by differential relations like
(19)-(20), and the fields $\epsilon$, $\beta$, $\bm{E}$,
$\bm{\mathfrak{B}}$ by the multiplier $\varkappa$. So, the fields
$V^\mu$, $U^\mu$ are connected with all potentials in the same way
as the fields $\epsilon$, $\beta$, $\bm{E}$, $\bm{\mathfrak{B}}$.

Substitution of the expressions (19)-(22) into Eqs. (15)-(18) yields
equations (7) and (13) for the potentials.

The Eqs. (15)-(18) are the Euler equations for the Lagrangian
\begin{equation} L=L_b+L_i,
\end{equation}
\[L_b=\frac{1}{2\zeta}(-\frac{1}{2}F_{\alpha\beta}F^{\alpha\beta}-
\epsilon^2+\beta^2+ V_\alpha V^\alpha-U_\alpha U^\alpha)=\]
\begin{equation}
=\frac{1}{2\zeta}(E^2-\mathfrak{B}^2-\epsilon^2+ \beta^2+V_0^2-V^2
-U_0^2+U^2),
\end{equation}
\begin{equation}
L_i=- j_e^\alpha X_\alpha+ j_m^\alpha Y_\alpha,
\end{equation}
if we consider the potentials $X^\mu$ and $Y^\mu$  as
independent variables. The interaction Lagrangian $L_i$ describes
the minimal interaction of the  vector and pseudovector bosons $X^\mu$,
$Y^\mu$ with the currents $j_e^\alpha$, $j_m^\alpha$ that produce
them. Even if these currents are not conserved the propagators of
bosons $X^\mu$ and $Y^\mu$ have the form (8).

The non-conserved currents are met in the study of the weak
interaction.   Let us use an example of  the  weak interaction of
the charged currents in order to show  the usefulness of the Eqs.
(5), (11), (15)-(22) to describe the vector and pseudovector bosons.

In the Weinberg-Salam electroweak model the charged current interaction
is given by  the Lagrangian \cite{com}:
\begin{equation}
L_i\!=-\frac{g}{2\sqrt{2}}\bar{e}\gamma^\mu(1-\gamma^5)W_\mu^{-}\nu-
\frac{g}{2\sqrt{2}}\bar{\nu}\gamma^\mu(1-\gamma^5)W_\mu^{+}e.
\end{equation}
Here, $\gamma^5=i\gamma^0\gamma^1\gamma^2\gamma^3$. The potentials
$W^{+}_\mu$, $W^{-}_\mu$ describe the charged bosons. Let us show
that the Lagrangian (30) coincides with (29), if we choose as the
currents $j^\mu_e$, $j^\mu_m$  the following combinations in Eq.
(29):
\begin{equation}
j_e^\mu=\frac{g}{\sqrt{2}}(\bar{\nu}\gamma^\mu
e+\bar{e}\gamma^\mu\nu),
\end{equation}
\begin{equation}
j_m^\mu=i\frac{g}{\sqrt{2}} (\bar{\nu}\gamma^\mu\gamma^5e-
\bar{e}\gamma^\mu\gamma^5\nu).
\end{equation}
Really, the Lagrangian (29) for the currents (31)-(32) is the
following:
\[L_i=- j_e^\mu X_\mu+ j_m^\mu Y_\mu=\]
\[-\frac{g}{\sqrt{2}}\Bigl[(\bar{e}\gamma^\mu\nu+\bar{\nu}\gamma^\mu
e) X_\mu+
i(\bar{e}\gamma^\mu\gamma^5\nu-\bar{\nu}\gamma^\mu\gamma^5e)
Y_\mu\Bigr]\!\!=\]
\[-\frac{g}{\sqrt{2}}\Bigl[\bar{e}\gamma^\mu(X_\mu+i\gamma^5Y_\mu)\nu
+\bar{\nu}\gamma^\mu(X_\mu-i\gamma^5Y_\mu)e\Bigr]=\]
\[-\frac{ g}{2\sqrt{2}}\bar{e}\gamma^\mu
\bigl[(1+\gamma^5)+(1-\gamma^5)\bigr]
(X_\mu+i\gamma^5Y_\mu)\nu-\]
\[-\frac{ g}{2\sqrt{2}}
\bar{\nu}\gamma^\mu\bigl[(1+\gamma^5)+(1-\gamma^5)\bigr]
(X_\mu-i\gamma^5Y_\mu)e=\]
\[-\frac{g}{2\sqrt{2}}\bar{e}\gamma^\mu\bigl[(1+\gamma^5)(X_\mu+iY_\mu)+
(1-\gamma^5)(X_\mu-iY_\mu)\bigr]\nu-\]
\begin{equation}
-\frac{g}{2\sqrt{2}}\bar{\nu}\gamma^\mu\bigl[(1+\gamma^5)(X_\mu-iY_\mu)+
(1-\gamma^5)(X_\mu+iY_\mu)\bigr]e.
\end{equation}

Due to the fact that only the left neutrino exists, all terms with
the factor $1+\gamma^5$ are zeros, and  Eq. (33) becomes:
\begin{equation}
L_i=-\frac{g}{2\sqrt{2}}\bar{e}\gamma^\mu(1-\gamma^5)W_\mu^{-}\nu-
\frac{g}{2\sqrt{2}}\bar{\nu}\gamma^\mu(1-\gamma^5)W_\mu^{+}e.
\end{equation}
In this expression
\begin{equation}
W_\mu^{-}=X_\mu-iY_\mu,\hspace{7mm} W_\mu^{+}=X_\mu+iY_\mu.
\end{equation}

The Lagrangian (34) coincides with (30).
The potentials $W_\mu^{+}$ and $W_\mu^{-}$ are given
by the combination of the  Eqs. (7) and  (13) :
\begin{equation}
\square\genfrac{}{}{0pt}{}{W^{+\mu}}{W^{-\mu}}+
\varkappa^2\genfrac{}{}{0pt}{}{W^{+\mu}}{W^{-\mu}}\!=\!
\genfrac{}{}{0pt}{}{\zeta(j^\mu_e+ij^\mu_m)\!=\!
\zeta\frac{g}{\sqrt{2}}\bar{\nu}\gamma^\mu(1-\gamma^5)e}
{\zeta(j^\mu_e-ij^\mu_m)\!=\!
\zeta\frac{g}{\sqrt{2}}\bar{e}\gamma^\mu(1-\gamma^5)\nu}.
\end{equation}
We can introduce fields:
\begin{equation}
\bm{E}^\pm= -\frac{1}{c}\frac{\partial\bm{W}^\pm}{\partial t}-\nabla
W_0^\pm\mp i\nabla\times\bm{W}^\pm,
\end{equation}
\begin{equation}
\epsilon^\pm= -\frac{1}{c}\frac{\partial W_0^\pm}{\partial
t}+\nabla\cdot\bm{W}^\pm,\hspace{7mm} V_\mu^\pm=-\varkappa
W_\mu^\pm.
\end{equation}
These fields are determined from the equations
\begin{equation}
\frac{1}{c}\frac{\partial \epsilon^\pm}{\partial
t}+\nabla\cdot\bm{E}^\pm-\varkappa V_0^\pm=\zeta c\rho^\pm,
\end{equation}
\begin{equation}
\frac{1}{c}\frac{\partial\bm{E}^\pm}{\partial t}\mp
i\nabla\times\bm{E}^\pm+\nabla \epsilon^\pm+ \varkappa\bm{V}^\pm=
-\zeta\bm{j}^\pm,
\end{equation}
where $\rho^\pm=\rho_e\pm i\rho_m$, $\bm{j}^\pm=\bm{j}_e\pm
i\bm{j}_m$. As  can be  seen the superpositions of the vector and
pseudovector currents $j_e^\mu\pm ij_m^\mu$ create the potentials
$W^{+\mu}$ and $W^{-\mu}$. Each superposition creates those
potential, with which it interacts in Eq. (34). Though the currents
(31) and (32) are not conserved, the propagators of the bosons
$W^{+\mu}$ and $W^{-\mu}$ still have the same form (8).

It is also follows from  Eq. (35)  that the charged bosons can be
considered as the superpositions of two states.   These states are
determined by the modified Maxwell's equations (15)-(18) with the
currents (31) and (32). One of them is described by the vector
potential $X^\mu$, the other by the pseudovector potential $Y^\mu$.
These potentials interact minimally with the currents (31) and (32),
respectively.

Thus, the bosons, the leptons and the weak interaction of their
charged currents can be described by the Lagrangian
\begin{equation}
L=L_b+\hbar cL_l+L_i,
\end{equation}
where
\[L_l=\frac{i}{2}(\bar{e}\gamma^\mu\frac{\partial e}{\partial
x^\mu}-\frac{\partial\bar{e}}{\partial x^\mu}\gamma^\mu
e)-\frac{m_ec}{\hbar}\bar{e}e+\]
\begin{equation}
+\frac{i}{2}(\bar{\nu}\gamma^\mu\frac{\partial \nu}{\partial
x^\mu}-\frac{\partial\bar{\nu}}{\partial x^\mu}\gamma^\mu\nu).
\end{equation}
It should be pointed out that the interaction Lagrangian in (41) can
be used as (29), which is similar to the electrodynamics interaction
Lagrangian. The boson propagator is different from the photon
propagator by the mass term in the denominator only. That is why
such interaction is renormalized. Because the Lagrangians (29) and
(30) coincide, the results obtained with  the Lagrangian (29) and
obtained with  the Weinberg-Salam model Lagrangian (30) coincide
also.

Attention should be paid to the fact that the pseudovector current
$j_m^\mu$ plays the role of the magnetic current in the modified
Maxwell's equations (15)-(18). This leeds our attention back  to the
Dirac hypothesis about the magnetic monopole \cite{d}. Dirac
introduced the magnetic current in analogy to the electric one ---
as a product of a magnetic charge and a velocity of a particle.
Above, the magnetic current is introduced in a different way --- as
a product of the electric charge (the electric charge $e$ is
connected with the charge $g$, $g=e/\sin\theta_W$) and the
pseudovector
$i(\bar{\nu}\gamma^\mu\gamma^5e-\bar{e}\gamma^\mu\gamma^5\nu)$. In
such a way, the magnetic monopole current is introduced without the
magnetic monopole. The last definition does not require the
introduction of  magnetic monopole at all.

We do not touch on here other aspects of the weak interaction, as the
aim of the article is to show the usefulness of use of the equations
(5) and (11) instead of the Proca equation for the description of
the vector and pseudovector bosons in the case of  non-conserved
current interactions.

\subsection*{Acknowledgments}
The author would like to thank Prof. Lev B.I., Prof. Lukyanets S.P.
and Prof. Zhmudskyy O.O. for stimulating discussions.

\end{document}